\documentclass[nature
,showpacs
,superscriptaddress
]
{revtex4-1}
\usepackage{graphicx,subfigure,bbm}

\usepackage{amsmath}
\usepackage{amsthm}
\usepackage{amssymb}
\usepackage{array,bm}
\usepackage{graphicx}
\usepackage{fancyhdr}
\usepackage{color}
\usepackage{extarrows}
\usepackage{enumerate}
\usepackage{epstopdf}
\usepackage[colorlinks,
            linkcolor=black,
            citecolor=black,
            urlcolor=blue
            ]{hyperref}
\usepackage{natbib}
\usepackage[latin1]{inputenc}
\usepackage{tikz}
\usepackage[displaymath, mathlines]{lineno}
\usetikzlibrary{shapes,arrows}

\tikzstyle{decision} = [diamond, draw, fill=blue!20,
    text width=4.5em, text badly centered, node distance=3cm, inner sep=0pt]
\tikzstyle{block} = [rectangle, draw, fill=blue!20,
    text width=10em, text centered, rounded corners, minimum height=4em]
\tikzstyle{line} = [draw, -latex']
\tikzstyle{cloud} = [rectangle, draw,fill=red!20, node distance=7cm,
    minimum height=4em]

\def\(({\left(}
\def\)){\right)}
\def\[[{\left[}
\def\]]{\right]}

\newcommand{\be}{\begin{equation}}
\newcommand{\ee}{\end{equation}}
\newcommand{\bea}{\begin{eqnarray}}
\newcommand{\eea}{\end{eqnarray}}

\DeclareMathAlphabet{\varmathbb}{U}{bbold}{m}{n}

\begin{document}
\title{Optimizing higher-order network topology for synchronization of coupled phase oscillators}

\author{Ying Tang}
\affiliation{International Academic Center of Complex Systems, Beijing Normal University, Zhuhai 519087, China}
\author{Dinghua Shi}
\affiliation{Department of Mathematics, College of Science, Shanghai University, Shanghai 200444, China}
\author{Linyuan L\"{u}}
\email{linyuan.lv@uestc.edu.cn}
\affiliation{Institute of Fundamental and Frontier Sciences, University of Electronic Science and Technology of China, Chengdu 610054, China}


\date{\today}
\begin{abstract}
\begin{center}
    \textbf{Abstract}
\end{center}
Networks in nature have complex interactions among agents. One significant phenomenon induced by interactions is synchronization of coupled agents, and the interactive network topology can be tuned to optimize synchronization. The previous studies showed that the optimized conventional network with pairwise interactions favors a homogeneous degree distribution of nodes when the interaction is undirected, and is always structurally asymmetric when the interaction is directed. However, the optimal control on synchronization for networks with prevailing higher-order interactions is less explored. Here, by considering the higher-order interactions in hypergraph and the Kuramoto model with 2-hyperlink interactions, we find that the network topology with optimized synchronizability may have distinct  properties. For the undirected interaction, optimized networks with 2-hyperlink interactions by simulated annealing tend to become homogeneous in the nodes' generalized degree, consistent with  1-hyperlink (pairwise) interactions. We further define the directed hyperlink, and rigorously demonstrate that for the directed interaction, the structural symmetry can be preserved in the optimally synchronizable network with 2-hyperlink interactions, in contrast to the conclusion for 1-hyperlink interactions. The results suggest that controlling the network topology of higher-order interactions leads to synchronization phenomena beyond pairwise interactions.
\end{abstract}

\maketitle

\section*{Introduction}


Complex interactions are ubiquitous in physical \cite{strogatz2001exploring,albert2002statistical}, biological \cite{barabasi2004network,bassett2006small}, and social systems \cite{amaral2000classes,lu2016vital}. The interactions form a complex network of the coupling agents. While a class of systems can be modeled by networks with pairwise interactions \cite{boccaletti2006complex,tang2017potential}, where nodes of the network are connected by links, higher-order interactions are prevailing in various systems, including the network of neurons \cite{reimann2017cliques}, the contagion network \cite{iacopini2019simplicial,Burgio2021Network}, and social networks \cite{alvarez2021evolutionary}. An emerging direction in network science started to uncover the significance of higher-order interactions \cite{shi2019totally,battiston2020networks,kovalenko2021growing,bick2020multi,young2021hypergraph,eriksson2021choosing}, which induce diverse phenomena beyond pairwise interactions.

Synchronization is one of the remarkable behaviors for coupled agents \cite{kuramoto1975self,strogatz2000kuramoto,RevModPhys.77.137,PhysRevX.9.011002}.
The previous studies revealed that synchronization depends on the topology of the network with pairwise interactions. For the undirected interaction, the more synchronizable networks of identical coupled agents tend to be homogeneous in the nodes' degree distribution \cite{shi2013searching,PhysRevLett.113.144101}.
For the directed interaction, except for the fully-connected network, the optimal network in synchronizability is always structurally asymmetric \cite{PhysRevLett.122.058301}.
Different from the pairwise interaction, higher-order interactions induce intriguing effects on synchronization \cite{PhysRevResearch.2.033410,skardal2020higher,gambuzza2021stability,zhang2020unified}. However, except for the specific network structure, e.g., star-clique topology \cite{PhysRevResearch.2.033410}, how does the network topology of higher-order interactions affect synchronization was seldom explored. Then, the question raises: for optimal synchronization, do the conclusions on the network topology with pairwise interaction \cite{shi2013searching,PhysRevLett.113.144101,PhysRevLett.122.058301}  still hold when higher-order interactions are present?

\begin{figure}[ht]
{\includegraphics[width=1\textwidth]{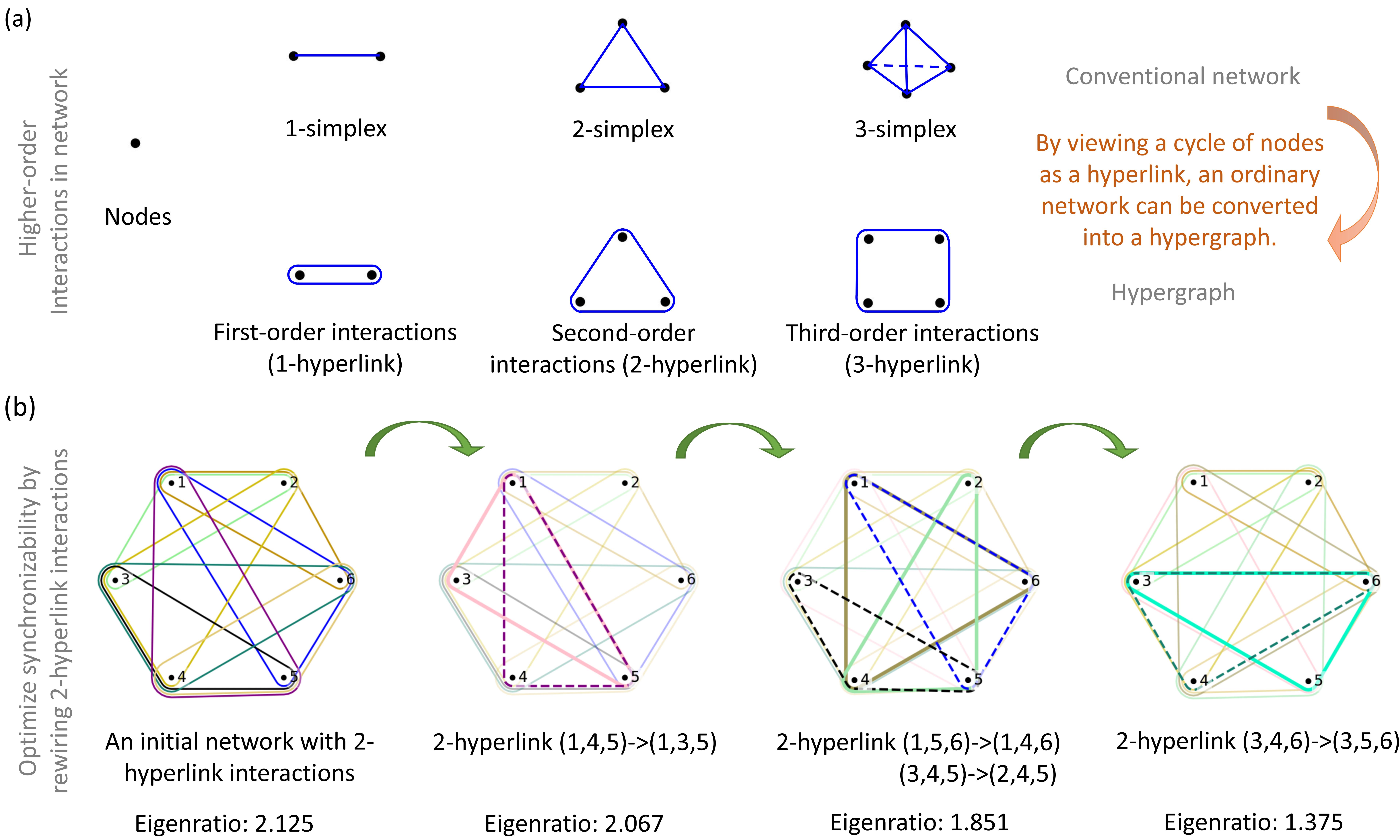}}
\caption{(Color online)
Schematic on optimizing synchronizability of the network with undirected higher-order interactions. The present higher-order interactions are from hyperlinks, similar as the simplex interactions \cite{PhysRevResearch.2.033410}. 
(a) Various orders of interactions, such as 1-hyperlink, 2-hyperlink, etc. 
The 2-hyperlink is an interaction among three nodes on hypergraphs, which is the present major focus. (b) An illustration on optimizing synchronizability for a network with undirected 2-hyperlink interactions. An initialized network with 6 nodes and 8 2-hyperlinks  is rewired to optimize its synchronizability, by minimizing the eigenratio of the generalized Laplacian. The colored triangles denote 2-hyperlink interactions, and are put on top of each other with shifting their positions for visualization. The rewiring process keeps the number of 2-hyperlinks constant, by deleting a few 2-hyperlinks and randomly adding the same number of 2-hyperlinks. At each step, the deleted triangles have dashed lines, and the added triangles have thicker solid lines. The procedure finds the network topology with smaller eigenratios and better synchronizability.  
}
\label{figure2New}
\end{figure}

In this paper, we investigate the effect of higher-order network topology on the phase synchronization of coupled oscillators on a type of hypergraph. By treating cycles on conventional networks as hyperlinks, a corresponding hypergraph can be obtained \cite{shi2019totally,fan2020characterizing}. Therein, higher-order interactions can be formulated as higher-order hyperlinks  \cite{battiston2020networks,skardal2020higher,gambuzza2021stability,zhang2020unified}, such as 1-hyperlink of two nodes  (first-order interaction, pairwise interaction), 2-hyperlink of three nodes  (second-order interaction), etc (Fig.~\ref{figure2New}).
The higher-order interactions from hyperlinks considered here is similar to the simplex interactions in \cite{PhysRevResearch.2.033410}, and are different from simplicial complexes \cite{PhysRevLett.124.218301} or the multilayer network \cite{boccaletti2014structure}. 

To analyze the network topology for optimal synchronization, 
we consider the Kuramoto-type coupling function for identical phase oscillators, focus on 2-hyperlink interactions, and search for the optimal network in synchronizability. Through analytical treatments and numerical estimations, we find that 2-hyperlink interactions can lead to distinct properties of the optimized networks compared with 1-hyperlink interactions. For the undirected interaction, we rewire 2-hyperlink interactions and use simulated annealing to optimize synchronizability by minimizing the eigenratios of generalized Laplacian matrices \cite{gambuzza2021stability,PhysRevResearch.2.033410}. Similar to  the conclusion for 1-hyperlink interactions \cite{shi2013searching,PhysRevLett.113.144101}, the optimized networks with 2-hyperlink interactions become homogeneous in the generalized nodes' degree. 

For the directed interaction, we provide an example of optimally synchronizable network with directed 2-hyperlink interactions that preserves structural symmetry (in the sense that each node has the same number and same type of higher-order interactions). We rigorously demonstrate that the optimally synchronizable network with higher-order interactions can have the symmetry, which is different from the result for 1-hyperlink interactions \cite{PhysRevLett.122.058301}. 
Still, the optimally synchronizable directed networks are found typically asymmetric by further numerical optimizations.
Overall, the present result uncovers that the properties of synchronizable networks with pairwise interactions may or may not hold for higher-order interactions, indicating that novel behaviors emerge in higher-order networks.

\section*{Results}

\textbf{Synchronizability of coupled phase oscillators with higher-order interactions.}\label{Sect2}
\textsl{The model and generalized Laplacians for higher-order interactions.}
We first present the formulation of the coupled oscillator system to study synchronization with higher-order interactions and the generalized higher-order Laplacians. 
We consider the Kuramoto-type model with the following set of ordinary differential equations for $N$ interacting phase oscillators ($N$ is also the network size) \cite{gambuzza2021stability}:
\begin{align}
\label{SynchronizationHighOrder}
\dot{\theta}_{i}&=f(\theta_{i})+K_{1}\sum_{l_{1}=1}^{N}a_{il_{1}}^{(1)}g^{(1)}(\theta_{i},\theta_{l_{1}})
+K_{2}\sum_{l_{1}=1}^{N}\sum_{l_{2}=1}^{N}a_{il_{1}l_{2}}^{(2)}g^{(2)}(\theta_{i},\theta_{l_{1}},\theta_{l_{2}})
\notag\\&\quad+\dots+K_{D}\sum_{l_{1}=1}^{N}\sum_{l_{2}=1}^{N}\dots\sum_{l_{D}=1}^{N}a_{il_{1}\dots l_{D}}^{(D)}g^{(D)}(\theta_{i},\theta_{l_{1}},\dots,\theta_{l_{D}}),
\end{align}
where $\theta_{i}\in[0,2\pi)$ is the one-dimensional state variable (the phase) for the $i$-th oscillator,
$f$ describes the local dynamics, $K_{d}$ ($d=1,2,\dots,D; D\leq N-1$)  are the coupling constants. Synchronization of the oscillators' phases is under consideration, which can be extended to state's synchronization by the master stability analysis \cite{gambuzza2021stability}.

For each order $d$,  $a_{il_{1}\dots l_{d}}^{(d)}$ are adjacency tensors. For example, the first-order interaction (1-hyperlink) has the conventional adjacency matrix: $a_{il_{1}}=1$ if the oscillators $(i,l_{1})$ have a pairwise interaction and $0$ otherwise; the second-order interaction (2-hyperlink) has $a_{il_{1}l_{2}}=1$ if the oscillators $(i,l_{1},l_{2})$ have a 2-hyperlink interaction and $0$ otherwise; etc. The interactions are undirected if the adjacency tensors are invariant under all permutation of  indices \cite{PhysRevResearch.2.033410}, and correspondingly are directed if such invariance does not hold, i.e., the adjacency tensors are variant under some permutation of  indices. 

The functions $g^{(d)}$ are coupling functions for synchronization, which is assumed to be non-invasive ($g^{(d)}(\theta,\theta,\dots,\theta)=0$ for $\forall d$)  \cite{gambuzza2021stability,zhang2020unified}. The  Kuramoto type of coupling functions \cite{kuramoto1975self,strogatz2000kuramoto,PhysRevLett.122.248301} have $g^{(1)}(\theta_{i},\theta_{l_{1}})=\sin(\theta_{l_{1}}-\theta_{i})$, $g^{(2)}(\theta_{i},\theta_{l_{1}},\theta_{l_{2}})=\sin(\theta_{l_{1}}+\theta_{l_{2}}-2\theta_{i})$, etc.  Then, the master stability equation \cite{PhysRevLett.80.2109,gambuzza2021stability} only depends on the adjacency tensors, i.e., generalized Laplacians, as the Jacobian terms in the master stability equation are constant \cite{PhysRevResearch.2.033410}. Besides, the present coupling constant $K_{d}$ can be denoted by the coefficients in \cite{PhysRevResearch.2.033410} as $K_{d}=\gamma_{d}/(d\langle K^{(d)}\rangle)$.


Based on the linearized equation in the master stability analysis (Methods), the generalized Laplacian matrix of the $d$-order interaction can be defined as \cite{PhysRevResearch.2.033410}:
\begin{align}
\label{GeneralizedDegAdj}
L_{ij}^{(d)}&=dk_{i}^{(d)}\delta_{ij}-k_{ij}^{(d)},\\
k_{i}^{(d)}&=\frac{1}{d!}\sum_{l_{1}=1}^{N}\dots\sum_{l_{d}=1}^{N}a_{i,l_{1},\dots,l_{d}}^{(d)},\\
k_{ij}^{(d)}&=\frac{1}{(d-1)!}\sum_{l_{2}=1}^{N}\dots\sum_{l_{d}=1}^{N}a_{i,j,l_{2},\dots,l_{d}}^{(d)},
\end{align}
where $k_{ij}^{(d)}$ is the generalized $d$-order degree between the nodes $i,j$, i.e., the number of $d$-hyperlink between $i,j$, and $k_{i}^{(d)}$
 is the generalized $d$-order degree of node $i$.
Note that for higher-order cases ($d>2$), the Laplacian here is different from another definition of the Laplacian, Eq.~\eqref{Laplacian0}. A detailed comparison on the various definitions of generalized Laplacians \cite{zhang2020unified,gambuzza2021stability,PhysRevResearch.2.033410} is in Methods. Further, we use the adjacency tensors to represent higher-order interactions, and employ higher-order Laplacians defined from adjacency tensors. Alternatively, higher-order interactions can be introduced by the boundary matrix acting on simplicial complexes \cite{PhysRevLett.124.218301,ghorbanchian2021higher}, which leads to a different way to define higher-order Laplacians.

To study the dependence of synchronizability on the network structure with higher-order interactions, we focus on the Kuramoto type of coupling function for the coupled oscillator. Then, the master stability equation, Eq.~\eqref{SynchronizationSecondOrderLinear2}, belongs to the case of Eq.~(15) in \cite{gambuzza2021stability}. As noted below Eq.~(15) in \cite{gambuzza2021stability}, the situation is conceptually equivalent to synchronization in networks with only pairwise interactions. The summation of higher-order Laplacians now plays the same role of the conventional Laplacian from pairwise interactions. Thus, synchronizability depends on generalized Laplacian matrices \cite{PhysRevLett.89.054101,PhysRevE.80.036204}, and can be characterized by the eigenvalues of Laplacian matrices \cite{nishikawa2010network,shi2013searching,PhysRevLett.122.058301,PhysRevResearch.2.033410}.

\textsl{The case with 2-hyperlink interactions.} In this study, we 
demonstrate the higher-order effect by studying 2-hyperlink interactions, i.e, the interaction among three nodes ($d=2$).
We further consider the identical oscillators, i.e., each oscillator has an identical frequency \cite{PhysRevLett.122.058301}, where the function $f(\theta)=\omega$ in Eq.~\eqref{SynchronizationSecondOrder}, with $\omega$ denoting the natural frequency of the oscillators. Then, 
the dynamical equation becomes (Methods):
\begin{align}
\label{SynchronizationSecondOrder}
\dot{\theta}_{i}=\omega
+K_{2}\sum_{l_{1}=1}^{N}\sum_{l_{2}=1}^{N}a_{il_{1}l_{2}}^{(2)}\sin(\theta_{l_{1}}+\theta_{l_{2}}-2\theta_{i}).
\end{align}
Its linearized synchronization dynamics is:
 \begin{align}
\label{SynchronizationSecondOrderLinear}
\delta\dot{\theta}_{i}=K_{2}\sum_{l_{1}=1}^{N}\sum_{l_{2}=1}^{N}a_{il_{1}l_{2}}^{(2)}(\delta\theta_{l_{1}}+\delta\theta_{l_{2}}-2\delta\theta_{i}).
\end{align}
Synchronizability is determined by the second-order Laplacian matrix:
\begin{align}
\label{SecondDegAdj}
L_{ij}^{(2)}&=2k_{i}^{(2)}\delta_{ij}-k_{ij}^{(2)},\\
k_{i}^{(2)}&=\frac{1}{2}\sum_{l_{1}=1}^{N}\sum_{l_{2}=1}^{N}a_{i,l_{1},l_{2}}^{(2)},\\
\label{GeneralizedDeg}
k_{ij}^{(2)}&=\sum_{l_{2}=1}^{N}a_{i,j,l_{2}}^{(2)}.
\end{align}
With 2-hyperlink interactions only, all the definitions of generalized Laplacians \cite{zhang2020unified,gambuzza2021stability,PhysRevResearch.2.033410} are the same (Methods).
In the next sections, we will separately study the optimized undirected and directed interactions for synchronizability.

\textbf{Optimized synchronizable networks with undirected 2-hyperlink interactions.}
\label{Sect3} \textsl{An example of a 6-node network with undirected 2-hyperlink interactions.}
\label{Sect2.3}
In this subsection, we give an example with its second-order adjacency tensor and generalized Laplacian to exemplify the network with 2-hyperlink interactions.  
Specifically, we consider the network with 6 nodes and 8 2-hyperlinks, and randomly initialize the 2-hyperlink interactions (Fig.~\ref{figure2New}). For example, the adjacency tensor is:
\begin{align}
\label{Example6NodeInitial}
a_{l_{1},l_{2},1}^{(2)}&=\left( \begin{array}{ccccccc}
0&0&0&0&0&0\\
0&0&1&0&0&1\\
0&1&0&0&0&0\\
0&0&0&0&1&0\\
0&0&0&1&0&1\\
0&1&0&0&1&0\\
\end{array} \right),\quad
a_{l_{1},l_{2},2}^{(2)}=\left( \begin{array}{ccccccc}
0&0&1&0&0&1\\
0&0&0&0&0&0\\
1&0&0&1&0&0\\
0&0&1&0&0&0\\
0&0&0&0&0&0\\
1&0&0&0&0&0\\
\end{array} \right),\quad
a_{l_{1},l_{2},3}^{(2)}=\left( \begin{array}{ccccccc}
0&1&0&0&0&0\\
1&0&0&1&0&0\\
0&0&0&0&0&0\\
0&1&0&0&1&1\\
0&0&0&1&0&0\\
0&0&0&1&0&0\\
\end{array} \right),
\notag\\
a_{l_{1},l_{2},4}^{(2)}&=\left( \begin{array}{ccccccc}
0&0&0&0&1&0\\
0&0&1&0&0&0\\
0&1&0&0&1&1\\
0&0&0&0&0&0\\
1&0&1&0&0&1\\
0&0&1&0&1&0\\
\end{array} \right),\quad
a_{l_{1},l_{2},5}^{(2)}=\left( \begin{array}{ccccccc}
0&0&0&1&0&1\\
0&0&0&0&0&0\\
0&0&0&1&0&0\\
1&0&1&0&0&1\\
0&0&0&0&0&0\\
1&0&0&1&0&0\\
\end{array} \right),\quad
a_{l_{1},l_{2},6}^{(2)}=\left( \begin{array}{ccccccc}
0&1&0&0&1&0\\
1&0&0&0&0&0\\
0&0&0&1&0&0\\
0&0&1&0&1&0\\
1&0&0&1&0&0\\
0&0&0&0&0&0\\
\end{array} \right),
\end{align}
where $l_{1},l_{2}$ are the first two indices. It has in total $8$ 2-hyperlinks connecting the nodes:
\begin{align}
(1,2,3),\quad(1,2,6),\quad(1,4,5),\quad(1,5,6),\quad(2,3,4),\quad(3,4,5),\quad(3,4,6),\quad(4,5,6).
\end{align}
Its generalized Laplacian matrix $L^{(2)}$ 
by Eq.~\eqref{SecondDegAdj} is: 
  \begin{align}
L^{(2)}=\left( \begin{array}{ccccccc}
8  &-2&-1&-1&-2&-2\\
-2&6  &-2&-1&0&-1\\
-1&-2&8  &-3&-1&-1\\
-1&-1&-3&10 &-3&-2\\
-2&0&-1&-3&8 &-2\\
-2&-1&-1&-2&-2&8
\end{array} \right).
\end{align}
The eigenvalues of generalized Laplacian matrix in ascending order are $0, 6.171, 8.167, 10.000, 10.549, 13.111$, and the eigenratio is $2.125$. We will provide networks with optimized synchronizability of this example below. 

\textsl{Optimizing synchronizability of undirected interactions by eigenratio of generalized Laplacian}. In this subsection, we present the framework to optimize synchronization of the network with  undirected interactions. We use the eigenratio of generalized Laplacians to quantify synchronization. 
Specifically, we calculate the eigenvalues of higher-order Laplacian matrices Eq.~\eqref{GeneralizedDegAdj} (or the sum of higher-order Laplacians if multiorder interactions are present \cite{PhysRevResearch.2.033410}), which can be arranged as $0=\lambda_{1}<\lambda_{2}\leq\dots\leq\lambda_{N}$. Note that the eigenvalues are all real, as generalized Laplacians are symmetric.  The smallest nonzero eigenvalue $\lambda_{2}$ is known as  the spectral gap. 
The eigenratio
\begin{align}
\label{eigenratio}
R=\lambda_{N}/\lambda_{2},
\end{align}
quantifies synchronizability \cite{gambuzza2021stability}. 
By diagonalizing the Laplacian matrix Eq.~\eqref{GeneralizedDegAdj}, we can get its eigenvalues and the eigenratio.

For the synchronized state of the system Eq.~\eqref{SynchronizationSecondOrder}, we consider its bounded and connected stability region, where synchronizability of coupled oscillators can be quantified in terms of the eigenratio Eq.~\eqref{eigenratio}. Then, synchronizability is optimized by minimizing the eigenratio Eq.~\eqref{eigenratio}, which was implemented mainly for networks with first-order interaction \cite{nishikawa2010network,shi2013searching,PhysRevLett.122.058301}. We optimize the networks with 2-hyperlink interactions for the linearized system Eq.~\eqref{SynchronizationSecondOrderLinear}, where eigenvalues of generalized Laplacians also determine synchronizability \cite{PhysRevResearch.2.033410}.

We make remarks about the following search for the optimally synchronizable networks. First,  we focus on using 2-hyperlink interactions to exemplify the effect of higher-order interactions in the optimized synchronizable network topology. The implementation can be extended to cases with higher-order interactions by a similar procedure.
Second, since we investigate the optimal network topology for synchronizability, we have considered an identical frequency for the oscillators. Under the case, the system has the global synchronization instead of the cluster synchronization \cite{pecora2014cluster,salova2021cluster}, and synchronizability is determined by the eigenratio.  


\textsl{The numerical protocol of optimizing networks with undirected 2-hyperlink interactions.}
\label{Sect2.1}
In this subsection, we  optimize synchronizability by numerically minimizing the eigenratio of the generalized Laplacian. 
We start with various randomly initialized networks, rewire second-order interactions with keeping a fixed number of 2-hyperlinks, and numerically search for optimal networks by simulated annealing  \cite{PhysRevLett.122.058301}  to minimize the eigenratio,  Eq.~\eqref{eigenratio} of the Laplacian matrix Eq.~\eqref{SecondDegAdj}. 

For the initialization, we randomly generate different networks with certain numbers of 2-hyperlinks. For $N$-node networks, there are $C_{N}^{3}=N(N-1)(N-2)/6$ combinations of three nodes, i.e., the number of 2-hyperlinks. To demonstrate the optimization procedure, we consider the network by first adding the 2-hyperlink interactions for the nodes $i,i+1,i+2$ $(i=1,\dots,N)$. It ensures that each node has at least one 2-hyperlink interaction, such that the network does not have isolated nodes. Then, we randomly add $N$ 2-hyperlink interactions to the network, such that each realization of optimization starts with these $N$ 2-hyperlinks generated differently. Note that the 5-node case by such an initialization is a fully-connected network and is already optimal in synchronization. Therefore, when rewiring undirected interactions, we choose the minimal network size as $6$.

Different ways of initialization can be implemented. For example, one may add  more 2-hyperlinks to get a smaller eigenratio and better synchronization for the network, which may no longer need to be optimized for synchronizability. Besides, the conventional networks with first-order interactions, such as the SF network or the Erdos-Renyi network \cite{PhysRevLett.113.144101}, could not be directly implemented for higher-order interactions. Thus, we have used the above initialized network by adding the 2-hyperlinks with certain rules.

When rewiring the network, we delete randomly  a few 2-hyperlink interactions, such as $\sim20\%$ of the existing 2-hyperlinks, and add the same number of 2-hyperlink interactions to three randomly chosen nodes which did not have an interaction before. For the 2-hyperlink interaction of the nodes $l_{1},l_{2},l_{3}$ to be deleted, the six elements of the adjacency tensor $a^{(2)}$, which have indices as all the permutation of $l_{1},l_{2},l_{3}$, are set to be zero: $a^{(2)}_{l_{i},l_{j},l_{k}}=0$ with $i,j,k$ being all the permutations of $1,2,3$. 
After each rewiring step, we calculate the eigenratio of the rewired network, and a Metropolis accept-reject step is used for the rewired network with smaller eigenratio \cite{PhysRevLett.122.058301}.

The optimization procedure is repeated until the eigenratio is smaller than a chosen target value, which is initially set as the smallest possible eigenratio $1$. However, the target eigenratio can not be too small, because otherwise it may not be reached due to the sparsity of 2-hyperlink interactions ($2N$) when the number of nodes increases. We thus increase the target eigenratio by $10\%$ if the rewiring step runs over $100$ times without reaching the current target eigenratio. This procedure automatically increases the target eigenratio, to reduce the computational time of searching for too small eigenratio that may not be achieved for the sparse large networks. It still ensures to optimize synchronizability by minimizing  the eigenratio. 

The network is rewired instead of simply deleting the 2-hyperlink interactions, because after deleting the 2-hyperlinks the eigenratio may be similar while all eigenvalues continue to become smaller. It eventually leads to an optimal network with much less 2-hyperlink interactions than the initial network. The optimization with only adding 2-hyperlink also gives less constrained network structures. Thus, we choose to rewire the network by adding the same number of 2-hyperlink interactions after the deletion, which preserves the total number of 2-hyperlink interactions. Different types of constraints can be employed in the optimization procedure, to search for the optimized network with desired properties. 

The above completes one realization of optimization, and $1000$ realizations are conducted with various configurations of initialized networks.
In total, we have two hyperparameters about the iteration. The first is the maximum number of reducing the eigenratio to the target value ($100$ times) before increasing the target value. The second is the number of realizations ($1000$), i.e., the number of different initialized networks. The computational time increases with these two hyperparameters, and also increases with the number of nodes, which may be hours when the size of the network is over $100$ on a personal desktop. One may decrease the number of independent realizations to reduce the computational time.

\textsl{Optimizing synchronizability of the 6-node network with undirected 2-hyperlink interactions.}
\label{Sect3.1}
In this subsection, we demonstrate the optimization procedure by the example above. 
In each step of the optimization, one or a few 2-hyperlinks are rewired to generate a network with a smaller eigenratio. An illustration is given in Fig.~\ref{figure2New}.  

After conducting the numerical optimization with preserving  6 nodes and 8 2-hyperlinks, the resultant optimized
network has the following second-order adjacency tensor:
\begin{align}
\label{Example6Node}
a_{l_{1},l_{2},1}^{(2)}&=\left( \begin{array}{ccccccc}
0&0&0&0&0&0\\
0&0&1&0&0&1\\
0&1&0&0&1&0\\
0&0&0&0&0&1\\
0&0&1&0&0&0\\
0&1&0&1&0&0\\
\end{array} \right),\quad
a_{l_{1},l_{2},2}^{(2)}=\left( \begin{array}{ccccccc}
0&0&1&0&0&1\\
0&0&0&0&0&0\\
1&0&0&1&0&0\\
0&0&1&0&1&0\\
0&0&0&1&0&0\\
1&0&0&0&0&0\\
\end{array} \right),\quad
a_{l_{1},l_{2},3}^{(2)}=\left( \begin{array}{ccccccc}
0&1&0&0&1&0\\
1&0&0&1&0&0\\
0&0&0&0&0&0\\
0&1&0&0&0&0\\
1&0&0&0&0&1\\
0&0&0&0&1&0\\
\end{array} \right),
\notag\\
a_{l_{1},l_{2},4}^{(2)}&=\left( \begin{array}{ccccccc}
0&0&0&0&0&1\\
0&0&1&0&1&0\\
0&1&0&0&0&0\\
0&0&0&0&0&0\\
0&1&0&0&0&1\\
1&0&0&0&1&0\\
\end{array} \right),\quad
a_{l_{1},l_{2},5}^{(2)}=\left( \begin{array}{ccccccc}
0&0&1&0&0&0\\
0&0&0&1&0&0\\
1&0&0&0&0&1\\
0&1&0&0&0&1\\
0&0&0&0&0&0\\
0&0&1&1&0&0\\
\end{array} \right),\quad
a_{l_{1},l_{2},6}^{(2)}=\left( \begin{array}{ccccccc}
0&1&0&1&0&0\\
1&0&0&0&0&0\\
0&0&0&0&1&0\\
1&0&0&0&1&0\\
0&0&1&1&0&0\\
0&0&0&0&0&0\\
\end{array} \right).
\end{align}
It has the following $8$ 2-hyperlinks connecting the nodes:
\begin{align}
(1,2,3),\quad(1,2,6),\quad(1,3,5),\quad(1,4,6),\quad(2,3,4),\quad(2,4,5),\quad(3,5,6),\quad(4,5,6).
\end{align}
The generalized Laplacian matrix  $L^{(2)}$ by Eq.~\eqref{SecondDegAdj} 
is:
  \begin{align}
L^{(2)}=\left( \begin{array}{ccccccc}
8  &-2&-2&-1&-1&-2\\
-2&8  &-2&-2&-1&-1\\
-2&-2&8  &-1&-2&-1\\
-1&-2&-1&8  &-2&-2\\
-1&-1&-2&-2&8 &-2\\
-2&-1&-1&-2&-2&8
\end{array} \right).
\end{align}
The eigenvalues in ascending order are $0, 8, 9, 9, 11, 11$, and the eigenratio is $1.375$.

We note that the optimized network may not be the best in synchronization until sufficient numerical search are conducted. However, with the present numerical optimization, this network is at least near to the optimal network in synchronizability as its eigenratio is close to $1$.


\begin{figure}[ht]
{\includegraphics[width=1\textwidth]{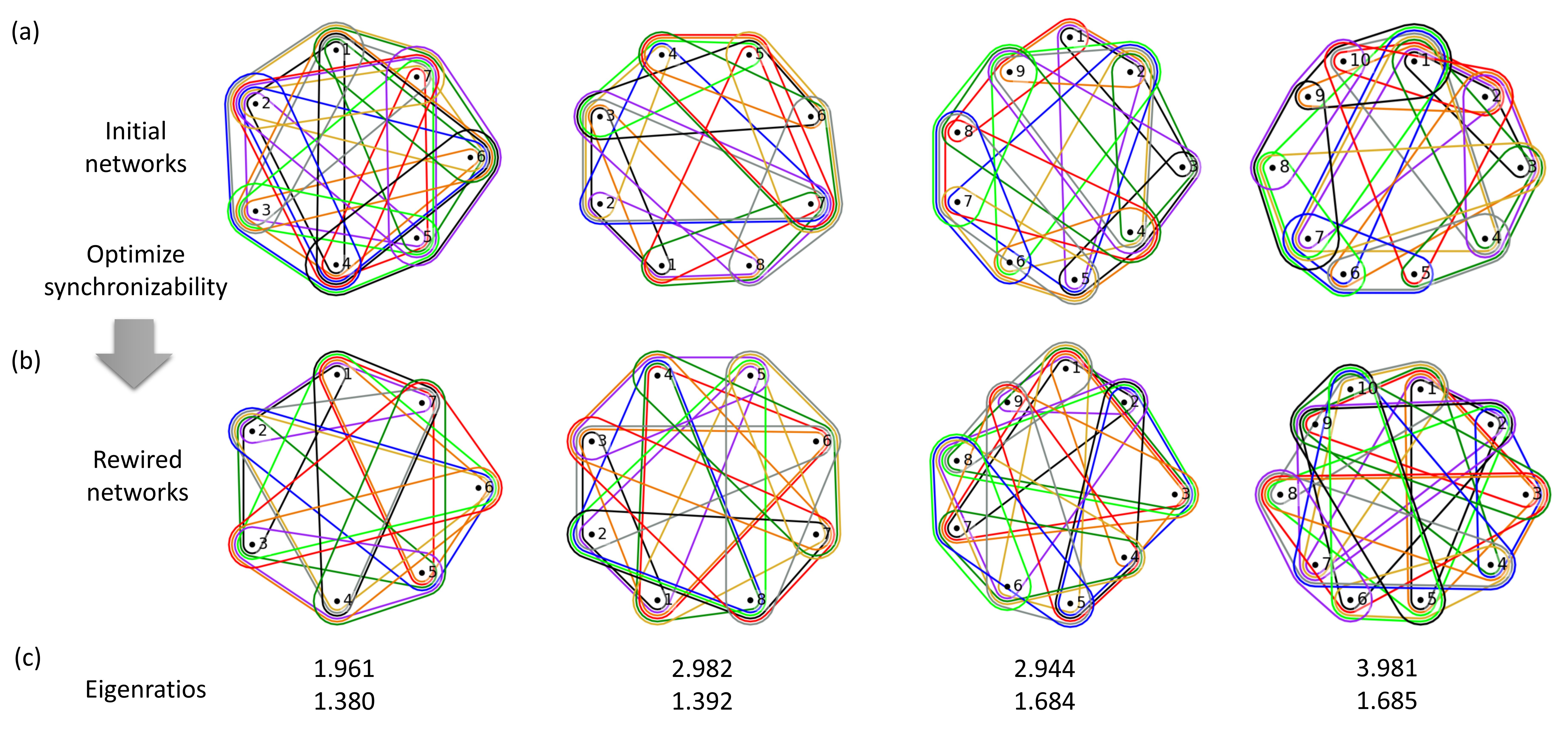}}
\caption{(Color online)
Examples of the network with 2-hyperlink interactions before and after the optimization on synchronizability. (a) The initial networks before the optimization. 
(b) The rewired network after the optimization.  We show the networks with small sizes, $N=7,8,9,10$, for better visualization on the 2-hyperlink interactions. (c) The eigenratios before and after the optimization. }
\label{figure2}
\end{figure}

\begin{figure}[ht]
{\includegraphics[width=0.8\textwidth]{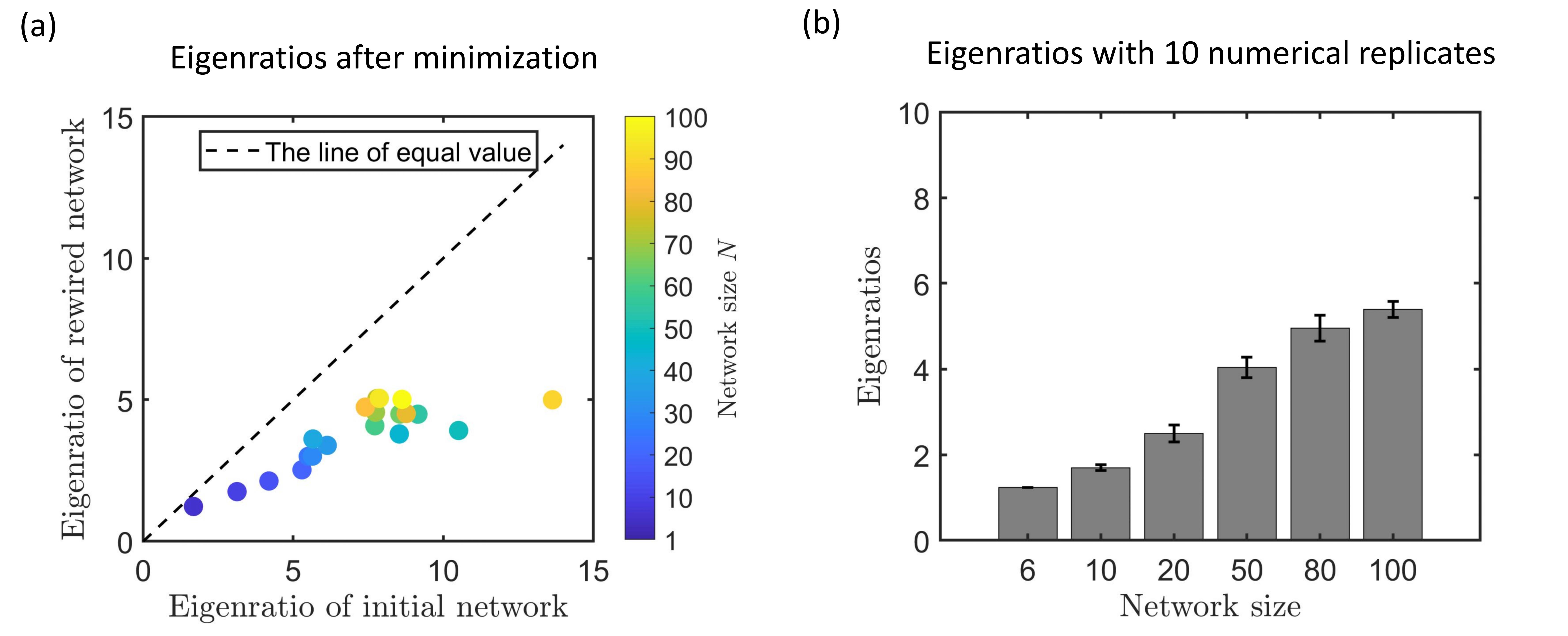}}
\caption{(Color online)
Eigenratios of networks with 2-hyperlink interactions before and after the optimization on synchronizability. (a) The eigenratios before and after the optimization, which shows that the optimizing procedure decreases eigenratios and enhances synchronizability. The network sizes include $6$ and those from $10$ to $100$ with a step size $5$, denoted by color. (b) The eigenratios of the optimized network with 2-hyperlink interactions for various size $6, 10, 20, 50, 80, 100$. 
The errorbar denotes the standard deviation of $10$ numerical replicates, and each replicate is the optimal from $1000$ different randomly initialized networks. 
}
\label{figure3}
\end{figure}

\begin{figure}[ht]
{\includegraphics[width=1\textwidth]{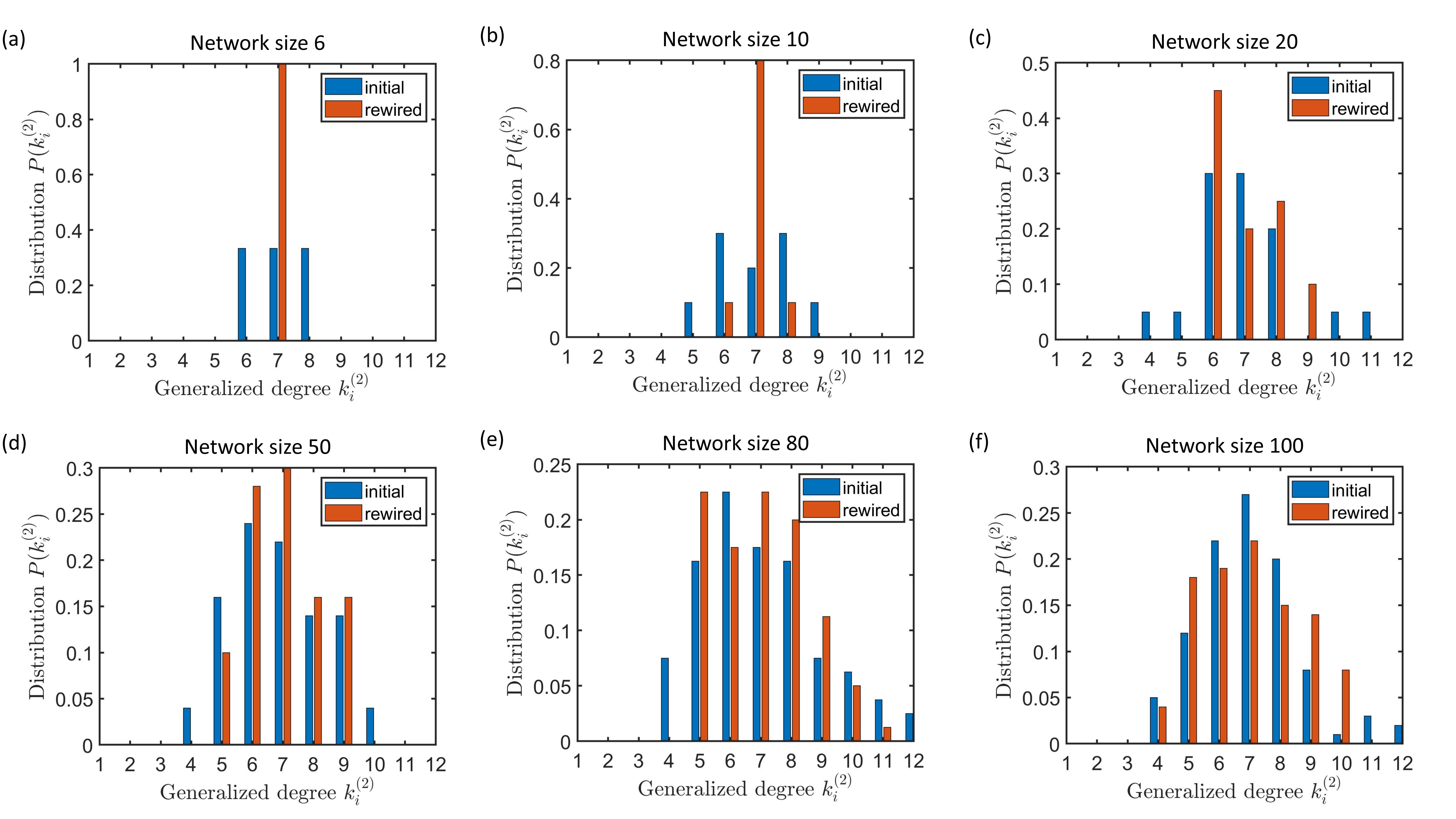}}
\caption{(Color online)
Statistical properties of the rewired undirected network after optimizing synchronizability. (a-f) The distribution of the generalized nodes' degree Eq.~\eqref{GeneralizedDeg} for networks with 2-hyperlink interactions under various sizes. The generalized nodes' degree distribution shows that the more synchronizable network tends to be homogeneous. 
It is consistent with 1-hyperlink interactions, where nodes also tend to be homogeneous in degree \cite{PhysRevLett.113.144101}. 
}
\label{figure4}
\end{figure}

\textsl{The optimized networks with various sizes.} In this subsection, we provide the optimization result for undirected networks with various sizes, as plotted in Fig.~\ref{figure2} and Fig.~\ref{figure3}. Examples of the initial and rewired network are given in Fig.~\ref{figure2}, with the number of nodes $N=7,8,9,10$. It demonstrates that the optimization procedure rewires 2-hyperlink interactions and reduces eigenratios. For illustration, we have shown networks with a small number of nodes. 

In Fig.~\ref{figure3}(a), the eigenratio before and after the optimization are provided, where the numbers of nodes include $6$ and those from $10$ to $100$ with a step size $5$. The optimized networks have smaller eigenratios after the optimization, showing better  synchronizability. With a fixed number of 2-hyperlink, the specific configuration of the optimized network does not dramatically affect the final eigenratio when conducting multiple numerical replicates (10) on the network structures, as shown by the errorbar in Fig.~\ref{figure3}.
The number of 2-hyperlinks is a crucial factor in determining synchronizability. The numbers of initialized 2-hyperlinks are $N+N$, which gives a sparse network and only allows a relatively large eigenratio after the optimization. The eigenratios are smaller when the number of 2-hyperlinks increases to improve synchronizability.  
At the same time, there is limited room to rewire the network for optimizing synchronizability if the number of 2-hyperlinks is abundant. 

We have further calculated the generalized degree $k_{i}^{(2)}$ in Eq.~\eqref{GeneralizedDeg} of each node, which quantifies the number of 2-hyperlinks participated by each node. The distribution of the generalized degree for the optimized networks with various sizes is in Fig.~\ref{figure4}. 
When considering a identical frequency distribution of  oscillators, the optimal network tends to be  more homogeneous in the nodes' degree for the network with first-order interactions \cite{shi2013searching,PhysRevLett.113.144101}. Similarly, the optimized network with 2-hyperlink interactions also become more homogeneous, as the nodes' degree distribution concentrates to fewer values of degree in Fig.~\ref{figure4}. 

\begin{figure}[ht]
{\includegraphics[width=0.7\textwidth]{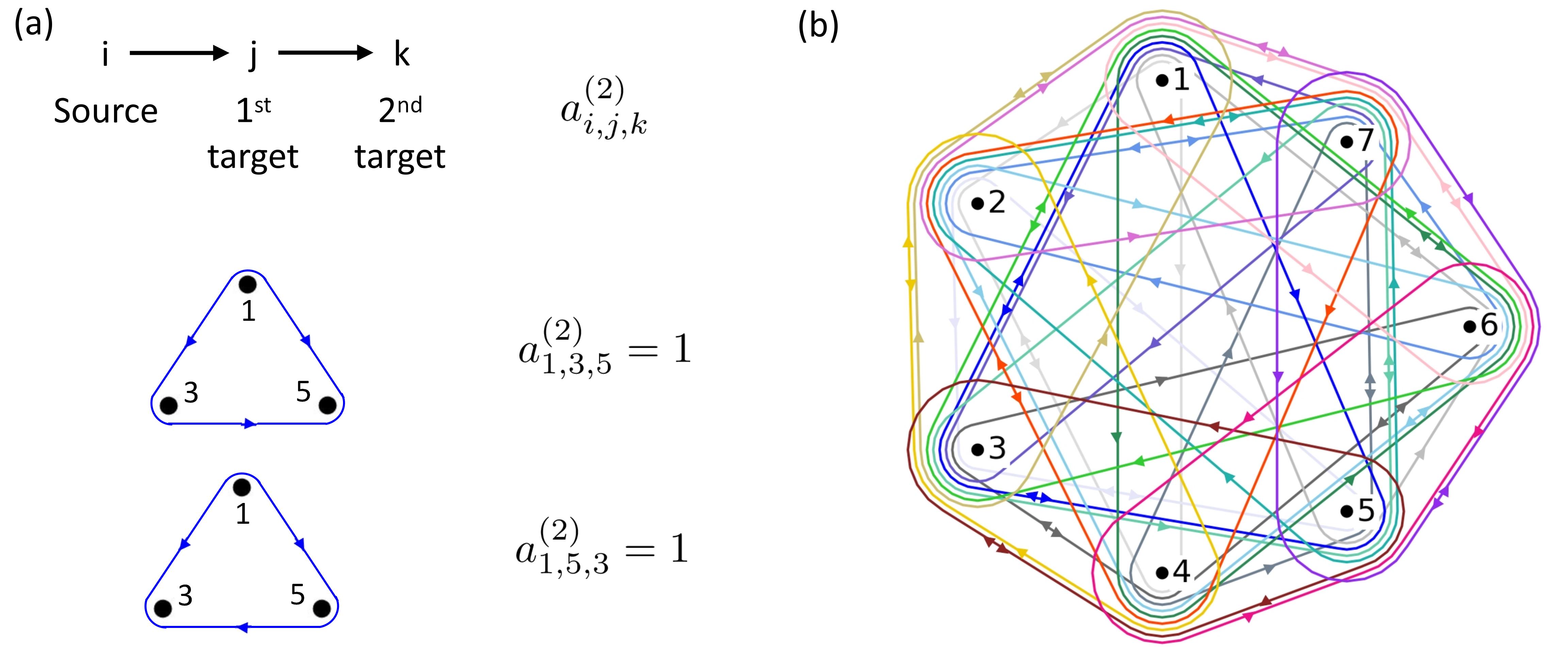}}
\caption{(Color online) An example of optimal network with directed 2-hyperlink interactions that preserves symmetry. (a) An illustration on the direction in directed 2-hyperlink interactions. The tensor $a^{(2)}_{i,j,k}$ has its subscript indices separately as source, 1st target, 2nd target nodes as in  \cite{PhysRevE.97.052303}. The directed 2-hyperlinks corresponding to $a^{(2)}_{1,3,5}=a^{(2)}_{1,5,3}=1$ are shown. (b) The $7$-node network with directed 2-hyperlink interactions. The 2-hyperlinks here are directed as given by the tensor in Eq.~\eqref{ExampleSymmetric2}: for example, the 2-hyperlink (1,3,5) has $a^{(2)}_{1,3,5}=a^{(2)}_{1,5,3}=1$ , and thus has the arrows $1\rightarrow3, 1\rightarrow5, 3\rightarrow5, 5\rightarrow3$ by the rule of setting the directed 2-hyperlink in the panel (a). 
This 7-node network shows that the structurally symmetric  network can be optimal in synchronization when 2-hyperlink interactions present. Differently, the optimal directed networks with pairwise interactions are always structurally asymmetric \cite{PhysRevLett.122.058301}.
}
\label{figure5}
\end{figure}

\textbf{Optimized synchronizable networks with directed 2-hyperlink interactions.}
\label{Sect4} In this section, we study synchronizability of the directed network.
We consider directed 2-hyperlink interactions to demonstrate the effect of directed higher-order interactions. 
The directed 2-hyperlinks are defined as that each permutation of three nodes leads to a distinct 2-hyperlink's direction: 
a 2-hyperlink interaction is ``directed'' once the tensor $a^{(2)}_{i,j,k}$, with $i,j,k$ being all the permutations of $1,2,3$, has one of its six elements to be nonidentical. 
Then, each $a^{(2)}_{i,j,k}$ with $i,j,k$ having a specific order can be regraded as a directed hyperedge, i.e., an ordered pair of disjoint subsets of vertices \cite{gallo1993directed}. The ordering $i,j,k$ specifies the direction, as illustrated in Fig.~\ref{figure5}. For example, with $a^{(2)}_{i,j,k}=1$, $i$ is a source node while $j,k$ are its target nodes, and $j$ is a source node while $k$ is its target node.

By using the adjacency tensors for directed interactions, the linearized synchronization dynamics is also given by Eq.~\eqref{SynchronizationSecondOrderLinear}, and synchronizability depends on eigenvalues of the generalized Laplacian matrix in Eq.~\eqref{SecondDegAdj}. The optimized directed network with only first-order interaction has been studied  \cite{PhysRevLett.122.058301}: it has been proved that the optimally synchronizable directed network is always structurally asymmetric (except for the fully-connected network). The proof is by establishing a contradiction that the structurally symmetric network can not be optimal in synchronizability. However, whether the conclusion holds for the network with higher-order interactions remains unknown.

For a higher-order directed network, it is regarded as structurally symmetric when each node has the same number and same type of higher-order interactions. Under the definition, we find that the symmetry may hold for the network with directed higher-order interactions, i.e., there are structurally symmetric networks which are optimal in synchronizability. Intuitively, higher-order interactions provide more capacity to reach optimal network design, enabling optimal synchronized directed networks to preserve structural symmetry.

\textsl{An example of optimally synchronizable network with directed 2-hyperlink interactions that preserves structural symmetry.} First, we demonstrate that a network with $N=7$ nodes and a symmetric structure can be optimal in synchronizability. As illustrated in Fig.~\ref{figure5},
the network has the following $7\times3$ directed 2-hyperlinks:
\begin{align}
\label{DirectedTriangle}
&(1,2,4),\quad(2,3,5),\quad(3,4,6),\quad(4,5,7),\quad(5,6,1),\quad(6,7,2),\quad(7,1,3),
\notag\\&
(1,3,5),\quad(2,4,6),\quad(3,5,7),\quad(4,6,1),\quad(5,7,2),\quad(6,1,3),\quad(7,2,4),
\notag\\&
(1,6,7),\quad(2,7,1),\quad(3,1,2),\quad(4,2,3),\quad(5,3,4),\quad(6,4,5),\quad(7,5,6).
\end{align}
Importantly, here each 2-hyperlink is directed, and it has the first node as the source node \cite{PhysRevE.97.052303}. For each 2-hyperlink in Eq.~\eqref{DirectedTriangle}, the nonzero elements in the adjacency tensor are chosen as:
\begin{align}
\label{ExampleSymmetric1}
a_{i,j,k}^{(2)}=a_{i,k,j}^{(2)}=1,\quad\text{for the 2-hyperlink}\quad (i,j,k).
\end{align}
Thus, the nonzero elements in the adjacency tensor are:
\begin{align}
\label{ExampleSymmetric2}
&a^{(2)}_{1,2,4}=1,\quad a^{(2)}_{2,3,5}=1,\quad a^{(2)}_{3,4,6}=1,\quad a^{(2)}_{4,5,7}=1,\quad a^{(2)}_{5,6,1}=1,\quad a^{(2)}_{6,7,2}=1,\quad a^{(2)}_{7,1,3}=1,
\notag\\&
a^{(2)}_{1,4,2}=1,\quad a^{(2)}_{2,5,3}=1,\quad a^{(2)}_{3,6,4}=1,\quad a^{(2)}_{4,7,5}=1,\quad a^{(2)}_{5,1,6}=1,\quad a^{(2)}_{6,2,7}=1,\quad a^{(2)}_{7,3,1}=1,
\notag\\&
a^{(2)}_{1,3,5}=1,\quad a^{(2)}_{2,4,6}=1,\quad a^{(2)}_{3,5,7}=1,\quad a^{(2)}_{4,6,1}=1,\quad a^{(2)}_{5,7,2}=1,\quad a^{(2)}_{6,1,3}=1,\quad a^{(2)}_{7,2,4}=1,
\notag\\&
a^{(2)}_{1,5,3}=1,\quad a^{(2)}_{2,6,4}=1,\quad a^{(2)}_{3,7,5}=1,\quad a^{(2)}_{4,1,6}=1,\quad a^{(2)}_{5,2,7}=1,\quad a^{(2)}_{6,3,1}=1,\quad a^{(2)}_{7,4,2}=1,
\notag\\&
a^{(2)}_{1,6,7}=1,\quad a^{(2)}_{2,7,1}=1,\quad a^{(2)}_{3,1,2}=1,\quad a^{(2)}_{4,2,3}=1,\quad a^{(2)}_{5,3,4}=1,\quad a^{(2)}_{6,4,5}=1,\quad a^{(2)}_{7,5,6}=1,
\notag\\&
a^{(2)}_{1,7,6}=1,\quad a^{(2)}_{2,1,7}=1,\quad a^{(2)}_{3,2,1}=1,\quad a^{(2)}_{4,3,2}=1,\quad a^{(2)}_{5,4,3}=1,\quad a^{(2)}_{6,5,4}=1,\quad a^{(2)}_{7,6,5}=1.
\end{align}
Other elements in the tensor $a^{(2)}$ are zero.

From this adjacency tensor, its generalized Laplacian matrix by Eq.~\eqref{SecondDegAdj} is: 
\begin{align}
L^{(2)}=\left( \begin{array}{ccccccc}
6  &-1&-1&-1&-1&-1&-1\\
-1&6  &-1&-1&-1&-1&-1\\
-1&-1&6  &-1&-1&-1&-1\\
-1&-1&-1&6  &-1&-1&-1\\
-1&-1&-1&-1&6  &-1&-1\\
-1&-1&-1&-1&-1&6  &-1\\
-1&-1&-1&-1&-1&-1 &6\\
\end{array} \right).
\end{align}
The Laplacian has all nonzero eigenvalues identical as $7$ and eigenratio $1$.

This network with 2-hyperlink interactions has the structural symmetry, as each node has the same number and type of 2-hyperlink interactions: each line of the 2-hyperlinks in Eq.~\eqref{DirectedTriangle} belong to the same type of 2-hyperlink for the first node (source node). Therefore, it is an counterexample of the structurally symmetric network being optimal in synchronizability. Higher-order interactions make it possible to have optimally synchronizable network with symmetry. We expect that larger networks can have more symmetric optimal structure.

\textsl{The optimally synchronizable network with directed 2-hyperlink interactions can preserve structural symmetry.} We next provide useful properties of eigenvalues for the Laplacian matrix $L^{(2)}$ of the optimally synchronizable network. The unweighted higher-order network is most synchronizable when the real eigenratio (Eq.~\eqref{eigenratio2}) is the smallest. This condition implies that the eigenvalues of the optimally synchronizable network are all real, such that they can be ordered, even though the directed networks generally have complex eigenvalues \cite{nishikawa2010network}.
That is, the nonzero eigenvalues of the Laplacian matrix satisfy $\lambda_{2}=\lambda_{3}=\dots=\lambda_{N}$ when the network is most synchronizable. Further, the optimally synchronizable condition constrains these eigenvalues to be integers, which was derived for the case with first-order interaction \cite{nishikawa2010network}. We now extend this property of Laplacian's eigenvalues to the network with higher-order interactions.

First, for convenience we define $\lambda_{2}=\lambda_{3}=\dots=\lambda_{N}\doteq\bar{\lambda}>0$ with $\bar{\lambda}\doteq\sum_{i=2}^{N}\lambda_{i}/(N-1)$.
According to \cite{nishikawa2010network}, this condition ensures that all eigenvalues are real for the first-order case, which can be extend to the higher-order cases with using generalized Laplacian \cite{gambuzza2021stability}.
Second, following a similar procedure in  \cite{nishikawa2010network}, we found that the identical eigenvalues are integers, i.e. $\bar{\lambda}$ is an integer. Specifically, the characteristic polynomial of generalized Laplacian is:
$\det(L^{(2)}-\lambda I)=-\lambda(\lambda-\bar{\lambda})^{N-1}=(-1)^{N}\lambda^{N}+\dots-\bar{\lambda}^{N-1}\lambda$, where $I$ is the identity matrix. As $L^{(2)}$ has all integer entries, the coefficients of the characteristic polynomial should be all integers, and thus $C\doteq\bar{\lambda}^{N-1}$ is an integer. According to the definition of the generalized Laplacian $L^{(2)}$ in Eq.~\eqref{SecondDegAdj}, $tr(L^{(2)})=l=(N-1)\bar{\lambda}$, where $l$ denotes the number of elements $1$ in the adjacency tensor $a^{(2)}$. Then, $C=[l/(N-1)]^{N-1}\doteq(s/t)^{N-1}$, where integers $s$ and $t$ do not have common factors. For $Ct^{N-1}=s^{N-1}$, any prime factor $p$ of $C$ must be a factor of $s^{N-1}$ and consequently a factor of $s$. Then, $p^{N-1}$ needs to be a factor of $k$, because there are no common factors in $s$ and $t$. Therefore, any factor of $C$ has multiplicity $N-1$, giving $C=q^{N-1}$ with an integer $q$. This leads to $\bar{\lambda}^{N-1}=q^{N-1}$ and that $\bar{\lambda}$ is an integer.

By using the above properties of the eigenvalues to establish a contradiction, the authors \cite{PhysRevLett.122.058301} found that an optimally synchronizable network with first-order interaction (except for the fully-connected network) must be structurally asymmetric. Below, we provide an attempt to establish such a contraction for the network with directed 2-hyperlink interactions, and find that the contraction can no longer be established. 

We now try to establish a contradiction that the structurally symmetric network can not be optimal in synchronizability. On the one hand, in a symmetric network,  the nodes are structurally identical. It guarantees that the in-degrees and out-degrees from the 2-hyperlink interactions of all nodes must be equal, i.e., each two connected nodes have bidirectional tensors. Thus, $l$ needs to be divisible by $N$ 
if the network is symmetric. As an example, for the network with $4$ nodes and 2-hyperlink interactions, the condition of structurally identical requires each node to have the same degree and the same number of 2-hyperlinks. Then, this $4$-node case needs to have a fully-connected network, with all the $4$ 2-hyperlinks and $l=12$ divisible by $4$. 


On the other hand, when the network is optimal in synchronizability, the eigenvalues are integers and equal: $\lambda_{2}=\lambda_{3}=\dots=\lambda_{N}=\bar{\lambda}$, and $tr(L^{(2)})=l=(N-1)\bar{\lambda}$. These properties imply that $\bar{\lambda}=l/(N-1)$ and that $l$ must be divisible by $N-1$ if the network is most synchronizable. Therefore, $l$ must be divisible by $(N-1)N$ and then
\begin{align}
(N-1)N\leq l.
\end{align}
On the other hand, for the network with 2-hyperlink interactions,
\begin{align}
l\leq(N-2)(N-1)N.
\end{align}
The two conditions can be satisfied simultaneously when $(N-1)N\leq l\leq(N-2)(N-1)N$. It no longer constrains the network to be fully-connected as the case of the network with only first-order interactions \cite{PhysRevLett.122.058301}. Thus, the contradiction that the structurally symmetric network is not optimal in synchronizability can not be reached for the network with 2-hyperlink interactions.


\textsl{Optimizing synchronizability of directed network by the real eigenratio of generalized Laplacian.} Generalized Laplacians for higher-order interactions are typically not structurally symmetric, and  lead to complex eigenvalues. The eigenvalues of the second-order Laplacian $L^{(2)}$ can be listed in ascending order of their real parts: $\lambda_{1},\lambda_{2},\lambda_{3},\dots,\lambda_{N}$.
In the strong coupling regime, since the stability region of the fully-synchronized state is bounded and connected, synchronizability can be quantified by an eigenratio of the real part \cite{PhysRevLett.122.058301}: 
\begin{align}
\label{eigenratio2}
R=\Re(\lambda_{N})/\Re(\lambda_{2}),
\end{align}
where $\Re$ denotes the real part of the complex eigenvalues.

The network will be more synchronizable if this eigenratio is smaller. Note that here we can extend this property to the higher-order case, because generalized Laplacians play the same role on quantifying synchronizability when  Eq.~\eqref{SynchronizationSecondOrderLinear} has identical oscillators and specific coupling functions (see the discussion under Eq.~(15) of \cite{gambuzza2021stability}). We numerically optimize synchronizability of the directed network by minimizing Eq.~\eqref{eigenratio2}. 

\textsl{An example of optimized networks with directed 2-hyperlink interactions.}
We provide an example of optimized networks by the numerical optimization. The initial network has $6$ nodes and the following $8$ undirected 2-hyperlinks:
\begin{align}
\label{UndirectedTriangle2}
&(1,3,5),\quad (1,3,6),\quad  (1,4,5),\quad  (1,4,6) ,\quad (2,3,5) ,\quad (2,3,6),\quad  (2,4,5),\quad  (2,4,6).
\end{align}
The nonzero elements in the adjacency tensor are:
\begin{align}
\label{ExampleUndirected2}
&a^{(2)}_{1,3,5}=1,\quad a^{(2)}_{1,3,6}=1,\quad a^{(2)}_{1,4,5}=1,\quad a^{(2)}_{1,4,6}=1,\quad a^{(2)}_{2,3,5}=1,\quad a^{(2)}_{2,3,6}=1,\quad a^{(2)}_{2,4,5}=1
,\quad a^{(2)}_{2,4,6}=1,
\notag\\
&a^{(2)}_{1,5,3}=1,\quad a^{(2)}_{1,6,3}=1,\quad a^{(2)}_{1,5,4}=1,\quad a^{(2)}_{1,6,4}=1,\quad a^{(2)}_{2,5,3}=1,\quad a^{(2)}_{2,6,3}=1,\quad a^{(2)}_{2,5,4}=1
,\quad a^{(2)}_{2,6,4}=1,
\notag\\
&a^{(2)}_{3,1,5}=1,\quad a^{(2)}_{3,1,6}=1,\quad a^{(2)}_{4,1,5}=1,\quad a^{(2)}_{4,1,6}=1,\quad a^{(2)}_{3,2,5}=1,\quad a^{(2)}_{3,2,6}=1,\quad a^{(2)}_{4,2,5}=1
,\quad a^{(2)}_{4,2,6}=1,
\notag\\
&a^{(2)}_{3,5,1}=1,\quad a^{(2)}_{3,6,1}=1,\quad a^{(2)}_{4,5,1}=1,\quad a^{(2)}_{4,6,1}=1,\quad a^{(2)}_{3,5,2}=1,\quad a^{(2)}_{3,6,2}=1,\quad a^{(2)}_{4,5,2}=1
,\quad a^{(2)}_{4,6,2}=1,
\notag\\
&a^{(2)}_{5,1,3}=1,\quad a^{(2)}_{6,1,3}=1,\quad a^{(2)}_{5,1,4}=1,\quad a^{(2)}_{6,1,4}=1,\quad a^{(2)}_{5,2,3}=1,\quad a^{(2)}_{6,2,3}=1,\quad a^{(2)}_{5,2,4}=1
,\quad a^{(2)}_{6,2,4}=1,
\notag\\
&a^{(2)}_{5,3,1}=1,\quad a^{(2)}_{6,3,1}=1,\quad a^{(2)}_{5,4,1}=1,\quad a^{(2)}_{6,4,1}=1,\quad a^{(2)}_{5,3,2}=1,\quad a^{(2)}_{6,3,2}=1,\quad a^{(2)}_{5,4,2}=1
,\quad a^{(2)}_{6,4,2}=1.
\end{align}
Other elements of $a^{(2)}$ are zero.

By deleting directed 2-hyperlinks, various configurations of the optimal network with $R=1$ can be reached by the numerical optimization. For example, one optimized directed network from the numerical optimization has the nonzero elements in the adjacency tensor:
\begin{align}
&a^{(2)}_{6,3,1}=1,\quad a^{(2)}_{3,5,1}=1,\quad a^{(2)}_{4,6,1}=1,\quad a^{(2)}_{5,2,3}=1,\quad a^{(2)}_{1,5,3}=1,
\notag\\& a^{(2)}_{6,2,4}=1,\quad a^{(2)}_{3,2,5}=1
,\quad a^{(2)}_{2,3,5}=1,\quad a^{(2)}_{2,4,6}=1,\quad a^{(2)}_{1,4,5}=1,
\end{align}
with other elements of $a^{(2)}$ zero.
Its generalized Laplacian matrix by Eq.~\eqref{SecondDegAdj}
is:
  \begin{align}
L^{(2)}=\left( \begin{array}{ccccccc}
2  &0&0&-1&-1&0\\
0&2  &-1&-1&0&0\\
0&-1&2  &0&-1&0\\
0&0&0&1  &0&-1\\
0&-1&0&0&1 &0\\
0&-1&-1&0&0&2
\end{array} \right).
\end{align}
The eigenvalues in ascending order of real parts are $0, 2,2,2,2,2$, and the eigenratio is $R=1$.

\textsl{Optimized networks with directed 2-hyperlink interactions are generally asymmetric.}  Though the counterexample shows that the optimal network can be structurally symmetric  when higher-order interaction presents, it does not mean that optimal directed networks with higher-order interactions generally tend to be symmetric over asymmetric. We use simulated annealing to search for the optimized synchronizable network numerically. For the directed interaction,  we delete the directed 2-hyperlink interaction, rather than rewiring the interaction, such that the network can become directed.  
After removing directed higher-order interactions to be more synchronizable, the network typically becomes structurally asymmetric.

\begin{figure}[ht]
{\includegraphics[width=0.85\textwidth]{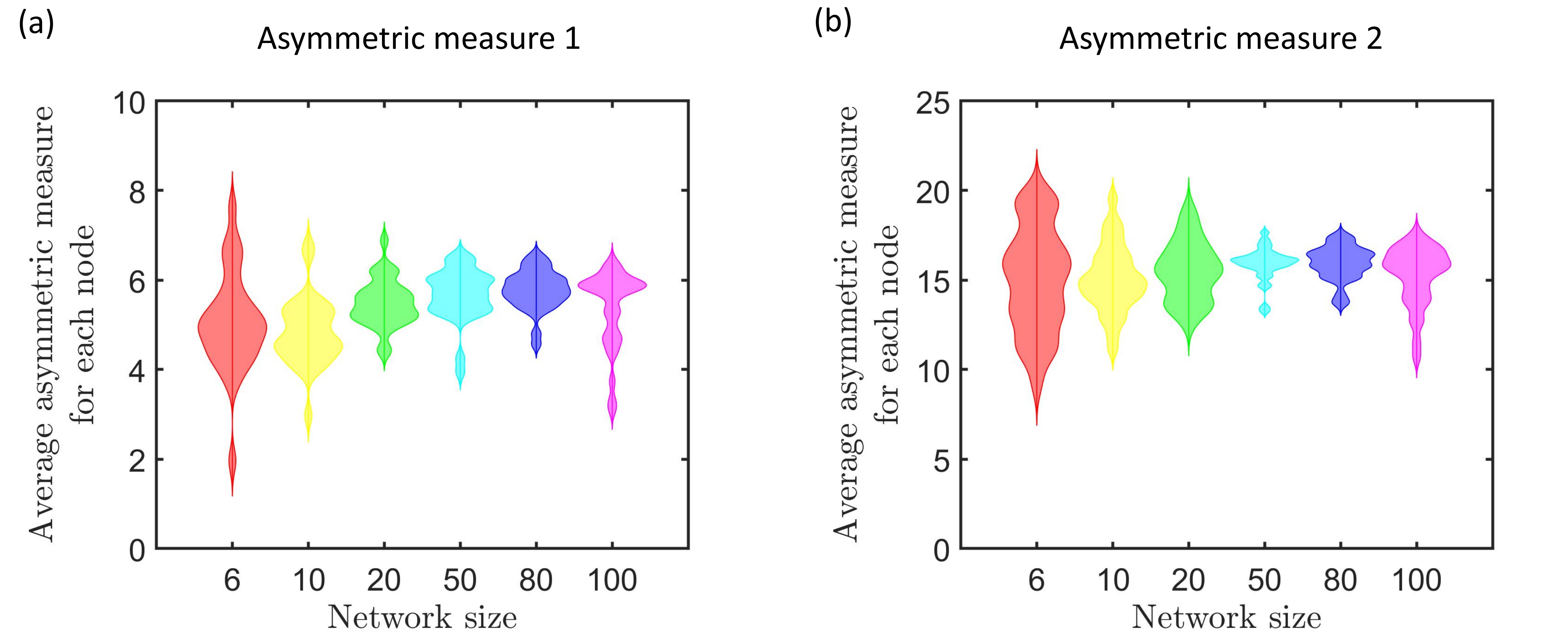}}
\caption{(Color online)
The optimized network with directed 2-hyperlink interactions are structurally asymmetric in general. Synchronizability of networks with directed 2-hyperlink interactions 
is optimized by minimizing Eq.~\eqref{eigenratio2} with simulated annealing. The directed 2-hyperlink  interactions are deleted, which allows to reach directed networks. (a,b) Violin distribution of the two asymmetric measures in Eq.~\eqref{measure1}, Eq.~\eqref{measure2} for the optimized directed networks with various size $6,10,20,50,80,100$, as specified by the color. They show that the optimized network are generally asymmetric, as most of the $30$ numerical replicates lead to nonzero asymmetric measures. Each numerical replicate has used the most synchronizable network optimized from $1000$ different randomly initialized networks.
}
\label{figure6}
\end{figure}

Specifically, we optimize synchronizability of the network by removing directed 2-hyperlink interactions, such that the network can become directed and asymmetric. By removing directed 2-hyperlinks, such as setting $a_{1,2,3}$, $a_{2,3,1}$ or $a_{3,1,2}$, etc to zero,  the network can have directed 2-hyperlink interactions instead of undirected interactions.
We employ simulated annealing for the optimization, where the input is a three-dimensional tensors for a network with 2-hyperlink interactions. We then calculate the generalized Laplacian matrix $L_{ij}^{(2)}$ for a given tensor $a_{l_{1},l_{2},l_{3}}$ by Eq.~\eqref{SecondDegAdj}, modify the network and calculate Eq.~\eqref{eigenratio2} after each step of modifying the network connection. We have used the same procedure to set the target value in the previous section, where the target value is increased by $10\%$ if the network modification runs over $100$ steps. Then, $1000$ realizations are conducted with different configurations of initialized networks.


We considered network sizes $6, 10, 20, 50, 80, 100$. After optimizing synchronizability by modifying the network, we use two quantities to measures the asymmetry of the optimized network. First,
we count the directed 2-hyperlink interaction of each node, which measures the structural asymmetry from each 2-hyperlink interaction for each node:
\begin{align}
\label{measure1}
A_{1}(l_{1})&=\sum_{l_{2}=1}^{N}\sum_{l_{3}=1}^{N}\Big|a_{l_{1},l_{2},l_{3}}^{(2)}-a_{l_{1},l_{3},l_{2}}^{(2)}\Big|,
\end{align}
where $A_{1}$ denotes the first asymmetry measure.
Second, we measure the asymmetry on the directed in-and-out interaction for three nodes of each 2-hyperlink interaction:
\begin{align}
\label{measure2}
A_{2}(l_{1})&=\sum_{l_{2}=1}^{N}\sum_{l_{3}=1}^{N}\Big|[a_{l_{2},l_{1},l_{3}}^{(2)}+a_{l_{2},l_{3},l_{1}}^{(2)}]+[a_{l_{3},l_{1},l_{2}}^{(2)}+a_{l_{3},l_{2},l_{1}}^{(2)}]
-2[a_{l_{1},l_{2},l_{3}}^{(2)}+a_{l_{1},l_{3},l_{2}}^{(2)}]\Big|,
\end{align}
where $A_{2}$ denotes the second asymmetry measure.

We estimate these asymmetry measures for each node, and then average them over all the nodes.  The two asymmetric measures $A_{1}, A_{2}$ of $30$ numerical replicates are plotted as violin distributions in Fig.~\ref{figure6}. They show that the optimized networks are generally asymmetric, because most numerical replicates generate the optimized network with nonzero asymmetric measures. Note that these measures may not quantify structural asymmetry of having 2-hyperlinks symmetrically in the network.

The phenomenon that asymmetry enhances synchronization is for the directed networks, which is not contradictory to optimal synchronization of fully-homogeneous undirected networks \cite{shi2013searching,shi2019totally}. Besides, after deleting 2-hyperlink interactions, the eigenvalues overall become smaller.  For the optimized network obtained above, we have chosen the minimum number of deletions to reach the target eigenratio. The procedure of deletion allows the network to be directed, and may not efficiently find the optimally synchronizable networks with structural symmetry. 

\section*{Conclusion}
\label{Sect5}
The recent studies \cite{alvarez2021evolutionary,battiston2020networks} revealed significant roles of higher-order interactions in network science. In the light of these studies, we investigate how higher-order interactions affect synchronization by optimizing network topology. Specifically, we have demonstrated the higher-order effect by focusing on the network topology of 2-hyperlink interactions, considering the phase synchronization of coupled oscillators, and employing the Kuramoto type of coupling function.
The cases with general coupling function can be studied by the master stability analysis \cite{PhysRevLett.80.2109,gambuzza2021stability,zhang2020unified}, which can determine the stability of the synchronized state. Under the case,  an interplay between the coupling function and the network topology needs to be analyzed. However, for a large class of coupled oscillator systems \cite{PhysRevLett.122.058301}, where the stability region is bounded and connected, synchronizability can be quantified by the eigenvalues of generalized Laplacian.

When oscillatory frequencies are heterogeneous, synchronizability depends on both network structure and oscillators' frequencies to be optimized. For pairwise interactions \cite{PhysRevLett.113.144101}, synchronization can be enhanced by a match between the heterogeneity of frequencies and network structure. For higher-order networks, one also needs to optimize the frequencies and the alignment function simultaneously. Here, we have focused on the network topology and considered identical oscillators. Besides, to study the network topology, we have treated the network as a single cluster,  rather than a network with a  sub-cluster coupled to other nodes \cite{PhysRevLett.122.058301}.

We have used adjacency tensors to encode higher-order interactions, which can be formulated as simplicial complexes or hypergraphs \cite{de2021phase}. We consider the case with pure 2-hyperlink interactions \cite{PhysRevResearch.2.033410}. 
Instead, simplicial complexes require collections of simplices \cite{battiston2020networks}. Extending the present result to simplicial complexes needs to include the various orders of simplices. 
To include multiorder interactions simultaneously depends on the coupling function. For general cases, there is an issue of diagonalizing the multiorder Laplacian matrices simultaneously \cite{zhang2020unified}, 
because the multiorder Laplacian matrices can not be directly added up due to the coupling function.
For the Kuramoto type of coupling function,  the Laplacian matrices can be added up \cite{PhysRevResearch.2.033410}, and then the multiorder eigenvalues 
determine the stability and quantify synchronizability. 
The numerical implementation is extendable to higher-order interactions by using generalized Laplacian matrices in Eq.~\eqref{GeneralizedDegAdj}.

When calculating eigenvalues of generalized Laplacian matrices, the analytical solvable cases are restricted to special networks \cite{PhysRevResearch.2.033410}, and numerical estimations are generally required. 
In the numerical implementation, the initialization on the network topology in Results 
ensures a sufficient number of 2-hyperlinks to be optimized. On the other hand, abundant 2-hyperlinks may cause an ineffectiveness of the optimization, 
because the network is already near to optimal synchronizability when the number of 2-hyperlinks is abundant. 
Different types of initialized networks can be used to further improve the search of the optimal network, with a cost of longer computational time.


To define the direction in higher-order networks, we have extended the definition of the directed network with pairwise interactions \cite{PhysRevLett.122.058301} to the 2-hyperlink case: the interactions are directed if the adjacency tensors are variant under the permutation of the indices. 
The direction of 2-hyperlink is assigned in a same way as that of the triangle in directed simplicial complexes \cite{PhysRevE.97.052303}. Moreover, the direction of hyperedge in  general hypergraphs needs to be carefully defined \cite{ducournau2014random}. To extend the present result of 2-hyperlink interactions to general hypergraphs, one needs to control hyperedge including their directions in the hypergraph.





In summary, our result demonstrates that higher-order interactions can lead to distinct properties of optimized synchronizability compared with the conventional network with pairwise interactions. For the undirected interaction, the more synchronizable networks tend to be homogeneous, consistent with pairwise interactions \cite{shi2013searching,PhysRevLett.113.144101}. For the directed interaction, the optimized synchronizable network is structurally asymmetric in general but can be symmetric, beyond the first-order case \cite{PhysRevLett.122.058301}. The optimization on synchronization of higher-order interactions may find uses in real networks, such as controlling the asynchronous state \cite{renart2010asynchronous} of brain networks with high-order structures  \cite{reimann2017cliques}.
Future work includes to investigate the optimal network design for chimera state \cite{PhysRevX.10.011044}, for the case with multiorder interactions of attractive and repulsive terms \cite{PhysRevResearch.2.033410}, and for general coupling functions.

{\small\section*{Methods}
\textbf{The master stability equation for the system of coupled oscillators.}
\label{Appendix1}
In this subsection, we present the method of master stability function \cite{PhysRevLett.80.2109}, which can be used to obtain the synchronization phase diagram for general coupled-oscillator systems Eq.~\eqref{SynchronizationHighOrder}. We consider the system with general coupling functions for the first and second-order interactions, as higher-order interactions can be treated similarly \cite{gambuzza2021stability}.

In order to study the stability of the synchronization solution, we consider small perturbations around the synchronous state, i.e., $\delta\theta_{i}=\theta_{i}-\theta^{\textbf{s}}_{i}$, where $\theta^{\textbf{s}}$ denotes the steady state. We then perform a linear stability analysis on the vector $\delta\theta=(\delta\theta_{1},\delta\theta_{2},\dots,\delta\theta_{N})$. The master stability equation \cite{PhysRevLett.80.2109} can be obtained as Eq.~(11) of \cite{gambuzza2021stability}:
\begin{align}
\label{MSE0}
\dot{\delta\theta}=[I_{N}\bigotimes JF-K_{1}L^{(1)}\bigotimes JG^{(1)}-K_{2}L^{(2)}\bigotimes JG^{(2)}]\delta\theta,
\end{align}
where $\bigotimes$ is the matrix Kronecker product, $I_{N}$ is the $n$-dimensional identity matrix and the Jacobian terms at the synchronized states are:
\begin{align}
JF&\doteq\frac{\partial f(\theta_{i})}{\partial\theta_{j}}\Big|(\theta^{\textbf{s}}),\\
JG^{(1)}&\doteq\frac{\partial g^{(1)}(\theta_{i},\theta_{j})}{\partial\theta_{j}}\Big|(\theta^{\textbf{s}},\theta^{\textbf{s}}),\\
JG^{(2)}&\doteq\frac{\partial g^{(2)}(\theta_{i},\theta_{l_{1}},\theta_{l_{2}})}{\partial\theta_{j}}\Big|(\theta^{\textbf{s}},\theta^{\textbf{s}},\theta^{\textbf{s}})+\frac{\partial g^{(2)}(\theta_{i},\theta_{l_{1}},\theta_{l_{2}})}{\partial\theta_{k}}\Big|(\theta^{\textbf{s}},\theta^{\textbf{s}},\theta^{\textbf{s}}).
\end{align}

After diagonalizing the first-order Laplacian matrix $L^{(1)}$ by a linear coordinate transformation $\delta\theta\rightarrow\delta\eta$, the master stability equation becomes Eq.~(13) of \cite{gambuzza2021stability}:
\begin{align}
\label{MSE}
\dot{\delta\eta}_{1}&=JF\delta\eta_{1},\\
\dot{\delta\eta}_{i}&=[JF-K_{1}\lambda_{i} JG^{(1)}]\delta\eta_{i}-K_{2}\sum_{j=2}^{N}\tilde{L}^{(2)}_{ij}\bigotimes JG^{(2)}\delta\eta_{i},
\end{align}
where $i=2,\dots,N$ denotes the  different modes transverse to the synchronization manifold, $\lambda_{i} $ are the eigenvalues of the Laplacian $L^{(1)}$, and $\tilde{L}^{(2)}$ is the transformed  second-order Laplacian $L^{(2)}$.

With the master stability equation Eq.~\eqref{MSE}, the master stability function $\Lambda_{max}$, i.e., the maximum transverse Lyapunov exponents (the modes except for the first stable mode), can be obtained by numerically tracking the norm of:
$\sqrt{\sum_{i=2}^{N}\sum_{j}^{m}[\eta_{i}^{(j)}]^{2}}$,
with $\eta_{i}\doteq(\eta_{i}^{(1)},\eta_{i}^{(2)},\dots,\eta_{i}^{(m)})$ solved from Eq.~\eqref{MSE} under given parameters. It is beyond solely using the eigenvalue of the Laplacian matrices for the Kuramoto-type coupling function. 

When  considering the Kuramoto-type coupling function and the linearized coupling terms \cite{shi2013searching}, the interaction term becomes variable-independent and eigenvalues of Laplacian matrix can quantify synchronization. Under the case, one could analyze eigenvalues of generalized Laplacian matrices Eq.~\eqref{GeneralizedDegAdj} to search for the optimally synchronizable network.  Without the linearization, the coupling function is variable-dependent. Then, we need to solve Eq.~\eqref{MSE} to quantify synchronizability, which prohibits to efficiently search for optimal network structure. We thus consider the Kuramoto-type coupling function \cite{PhysRevResearch.2.033410} in the main text. 

As for the linearization, in the strong coupling regime, the system Eq.~\eqref{SynchronizationSecondOrder} will converge to a synchronized state with $\theta_{i}\approx\theta_{j}$ for any $i,j$.
The linearized synchronization dynamics for the synchronized state is:
\begin{align}
\label{LinearSynchroSecondOrder}
\dot{\theta}_{i}=\omega+K_{1}\sum_{l_{1}=1}^{N}a_{il_{1}}^{(1)}(\theta_{l_{1}}-\theta_{i})
+K_{2}\sum_{l_{1}=1}^{N}\sum_{l_{2}=1}^{N}a_{il_{1}l_{2}}^{(2)}(\theta_{l_{1}}+\theta_{l_{2}}-2\theta_{i}).
\end{align}
The linearized equation reduces to the same form of Eq.~(2) in \cite{PhysRevResearch.2.033410}. 
We further use the rotating reference frame by $\theta_{i}=\theta_{i}-\omega t$, where the synchronized solution is $\theta_{i}(t)=0, (i=1,\dots,N)$. Then, the master stability equation for Eq.~\eqref{SynchronizationSecondOrder} is:
 \begin{align}
\label{SynchronizationSecondOrderLinear2}
\delta\dot{\theta}_{i}=K_{1}\sum_{l_{1}=1}^{N}a_{il_{1}}^{(1)}(\delta\theta_{l_{1}}-\delta\theta_{i})
+K_{2}\sum_{l_{1}=1}^{N}\sum_{l_{2}=1}^{N}a_{il_{1}l_{2}}^{(2)}(\delta\theta_{l_{1}}+\delta\theta_{l_{2}}-2\delta\theta_{i}),
\end{align}
which determines the linear stability of the synchronized state. This master stability equation is a case of Eq.~(15) in \cite{gambuzza2021stability}. It leads to Eq.~\eqref{SynchronizationSecondOrder} in the main text, where only the 2-hyperlink interaction is kept.

\textbf{Generalized Laplacians for higher-order interaction.}
\label{Appendix2}
In this section, we compare the different definitions on generalized Laplacians for higher-order interactions. In short, though their definitions for higher-order interactions in \cite{PhysRevResearch.2.033410,zhang2020unified,gambuzza2021stability} may differ, the first-order and second-order Laplacians are the same. As this paper focuses on the topology of 2-hyperlink interactions, the obtained results are valid for using any one of generalized Laplacians defined in \cite{PhysRevResearch.2.033410,zhang2020unified,gambuzza2021stability}. For convenience, we have used Eq.~\eqref{GeneralizedDegAdj0} as in \cite{PhysRevResearch.2.033410} when presenting the generalized higher-order Laplacians in the main text.

\begin{itemize}

 \item The Laplacian in \cite{PhysRevResearch.2.033410} is given by their Eqs.~(3,~4,~5):
\begin{align}
\label{GeneralizedDegAdj0}
L_{ij}^{(d)}&=dk_{i}^{(d)}\delta_{ij}-k_{ij}^{(d)},\\
k_{i}^{(d)}&=\frac{1}{d!}\sum_{l_{1}=1}^{N}\dots\sum_{l_{d}=1}^{N}a_{i,l_{1},\dots,l_{d}}^{(d)},\\
k_{ij}^{(d)}&=\frac{1}{(d-1)!}\sum_{l_{2}=1}^{N}\dots\sum_{l_{d}=1}^{N}a_{i,j,l_{2},\dots,l_{d}}^{(d)},
\end{align}
which is used in Eq.~\eqref{GeneralizedDegAdj} of the main text. We list it here for the comparison with other definitions of Laplacians.

\item The generalized Laplacians for higher-order interaction terms defined in \cite{gambuzza2021stability}  is by their Eq.~(28):
\begin{align}
\label{Laplacian0}
L_{ij}^{(d)}&=\left\{ \begin{array}{ccc}
&0&\quad i\neq j, \quad a_{ij}^{(1)}=0,\\
&-(d-1)!k_{ij}^{(d)}&\quad i\neq j, \quad a_{ij}^{(1)}=1,\\
&d! k_{i}^{(d)},&\quad i=j,\\
\end{array} \right.
\end{align}
where $k_{ij}^{(d)}$ also denotes the generalized $d$-order degree of the nodes $i,j$ and $k_{i}^{(d)}$
denotes the generalized $d$-order degree of node $i$.
Therein, their Eqs.~(24,~26) give:
 \begin{align}
k_{i}^{(d)}&=\frac{1}{d!}\sum_{l_{1}=1}^{N}\dots\sum_{l_{d}=1}^{N}a_{i,l_{1},\dots,l_{d}}^{(d)},\\
k_{ij}^{(d)}&=\frac{1}{(d-1)!}\sum_{l_{1}=1}^{N}\dots\sum_{l_{d-1}=1}^{N}a_{i,j,l_{1},\dots,l_{d-1}}^{(d)}.
 \end{align}
Here, $d$ specifies the order, which gives a $3$-dimensional tensor when $d=2$. 

Note that $k_{i}^{(d)}$, $k_{ij}^{(d)}$ here are the same as those in Eq.~\eqref{GeneralizedDegAdj0}, for example, the summations are the same. That is, $k_{i}^{(d)}$, $k_{ij}^{(d)}$ are the same in the literatures \cite{PhysRevResearch.2.033410,gambuzza2021stability}. However, their definitions of Laplacians are different, i.e., Eq.~\eqref{GeneralizedDegAdj0} and Eq.~\eqref{Laplacian0} are different. Still, for the first and second orders ($d=1, 2$), the two definitions of Laplacians reduce to the same.

  \item
The generalized Laplacian for higher-order interactions  used in \cite{zhang2020unified}  are:
\begin{align}
\label{Laplacian}
L_{ij}^{(1)}&=\delta_{ij}\sum_{k=1}^{N}a_{ik}^{(1)}-a_{ij}^{(1)},\\
L_{ij}^{(2)}&=-\sum_{k=1}^{N}a_{ijk}^{(2)}\quad (i\neq j),\quad L_{ii}^{(2)}=-\sum_{j\neq i}L_{ij}^{(2)},
\end{align}
to retain the zero row-sum property. It is consistent with that in \cite{gambuzza2021stability} for the second-order.
\end{itemize}

With using generalized Laplacians for the first-order and second-order cases, we obtain the optimized networks with 2-hyperlink interactions. 
\newline

\textbf{Data availability.} 
The authors declare that the data supporting the findings of this study are available within the paper. 

\textbf{Code availability}
The MATLAB code package is available at GitHub (https://github.com/jamestang23/Optimizing-higher-order-network-topology.git). All the simulations were done with MATLAB version R2019a.
}

\section*{Acknowledgments}
The work was supported by the National Natural Science Foundation of China (11622538, 61673150). L.L. also acknowledges the Science Strength Promotion Programme of University of Electronic Science and Technology of China.

\end{document}